\documentclass[conference]{IEEEtran}
\IEEEoverridecommandlockouts
\usepackage{cite}
\usepackage{amsmath,amssymb,amsfonts}
\usepackage{array}
\usepackage{graphicx}
\usepackage{textcomp}
\usepackage[dvipsnames,svgnames,x11names]{xcolor}
\def\BibTeX{{\rm B\kern-.05em{\sc i\kern-.025em b}\kern-.08em
    T\kern-.1667em\lower.7ex\hbox{E}\kern-.125emX}}
\newtheorem{myDef}{Definition}
\usepackage{subfigure}
\usepackage{algorithmic}
\usepackage{soul}
\usepackage[ruled]{algorithm2e}
\usepackage{makecell,multirow,diagbox}
\usepackage{booktabs}
\usepackage{multirow}
\usepackage{booktabs}
\usepackage{tabularx}
\begin{document}


\title{Generative Job Recommendations with Large Language Model}

\author{\IEEEauthorblockN{Zhi Zheng$^{1,2,\dagger}$, Zhaopeng Qiu$^{1,\dagger}$, Xiao Hu$^{1}$, Likang Wu$^{1,2}$, Hengshu Zhu$^{1*}$, Hui Xiong$^{3,4*}$
\thanks{$^\dagger$Equal Contribution.}
\thanks{$*$Corresponding authors.}}
\IEEEauthorblockA{
\textit{$^{1}$Career Science Lab, BOSS Zhipin}.\\
\textit{$^{2}$University of Science and Technology of China},\\
\textit{$^{3}$The Thrust of Artificial Intelligence, The Hong Kong University of Science and Technology}.\\
\textit{$^{4}$The Department of Computer Science and Engineering, The Hong Kong University of Science and Technology}.\\
zhengzhi97@mail.ustc.edu.cn, zhpengqiu@gmail.com, zhuhengshu@gmail.com, xionghui@ust.hk}}

\maketitle
\begin{abstract}
The rapid development of online recruitment services has encouraged the utilization of recommender systems to streamline the job seeking process. Predominantly, current job recommendations deploy either collaborative filtering or person-job matching strategies. However, these models tend to operate as ``black-box" systems and lack the capacity to offer explainable guidance to job seekers. Moreover, conventional matching-based recommendation methods are limited to retrieving and ranking existing jobs in the database, restricting their potential as comprehensive career AI advisors. To this end, here we present GIRL (GeneratIve job Recommendation based on Large language models), a novel approach inspired by recent advancements in the field of Large Language Models (LLMs). We initially employ a Supervised Fine-Tuning (SFT) strategy to instruct the LLM-based generator in crafting suitable Job Descriptions (JDs) based on the Curriculum Vitae (CV) of a job seeker. Moreover, we propose to train a model which can evaluate the matching degree between CVs and JDs as a reward model, and we use Proximal Policy Optimization (PPO)-based Reinforcement Learning (RL) method to further fine-tine the generator. This aligns the generator with recruiter feedback, tailoring the output to better meet employer preferences. 
In particular, GIRL serves as a job seeker-centric generative model, providing job suggestions without the need of a candidate set.
This capability also enhances the performance of existing job recommendation models by supplementing job seeking features with generated content. With extensive experiments on a large-scale real-world dataset, we demonstrate the substantial effectiveness of our approach. We believe that GIRL introduces a paradigm-shifting approach to job recommendation systems, fostering a more personalized and comprehensive job-seeking experience.

\end{abstract}


\section{Introduction}
Recent years have witnessed the rapid development of online recruitment. 
According to the report from The Insight Partners\footnote{https://www.theinsightpartners.com/reports/online-recruitment-market}, the global online recruitment market size is expected to grow from \$29.29 billion in 2021 to \$47.31 billion by 2028. 
For these platforms, the Recommendation Systems (RS) which can provide valuable assistance to job seekers by recommending suitable jobs, serve as their core components.
Alone this line, considerable research efforts have been made in building RS for online recruitment~\cite{DBLP:conf/cikm/LeHSZ0019,DBLP:conf/sigir/QinZXZJCX18,DBLP:journals/tmis/ZhuZXMXDL18}. 
Indeed, existing studies mainly follow the collaborative filtering~\cite{DBLP:conf/cikm/LeHSZ0019} or person-job matching paradigms~\cite{DBLP:conf/sigir/QinZXZJCX18,DBLP:journals/tmis/ZhuZXMXDL18}. 
However, in practical applications, these methods will encounter the following challenges. 
First, these methods primarily rely on end-to-end neural network models, which usually output the matching score directly given the information of a specific job seeker and a job.
Nevertheless, these models suffer from poor explainability in the black-box neural network computations, which reduce the user trust, especially in scenarios like job seeking that have a significant impact on individuals.
Second, most of the existing models are \emph{discriminative} model, which are limited to retrieving and ranking existing jobs in the database, restricting their potential as comprehensive career AI advisors, i.e., \emph{generating} a novel Job Description (JD) for a job seeker personally based on the Curriculum Vitae (CV).
Last but not least, the existence of the considerable semantic gap between CVs and JDs has resulted in the underwhelming performance of traditional methods.

\begin{figure*}[t]
\centering
\includegraphics[width=\textwidth]{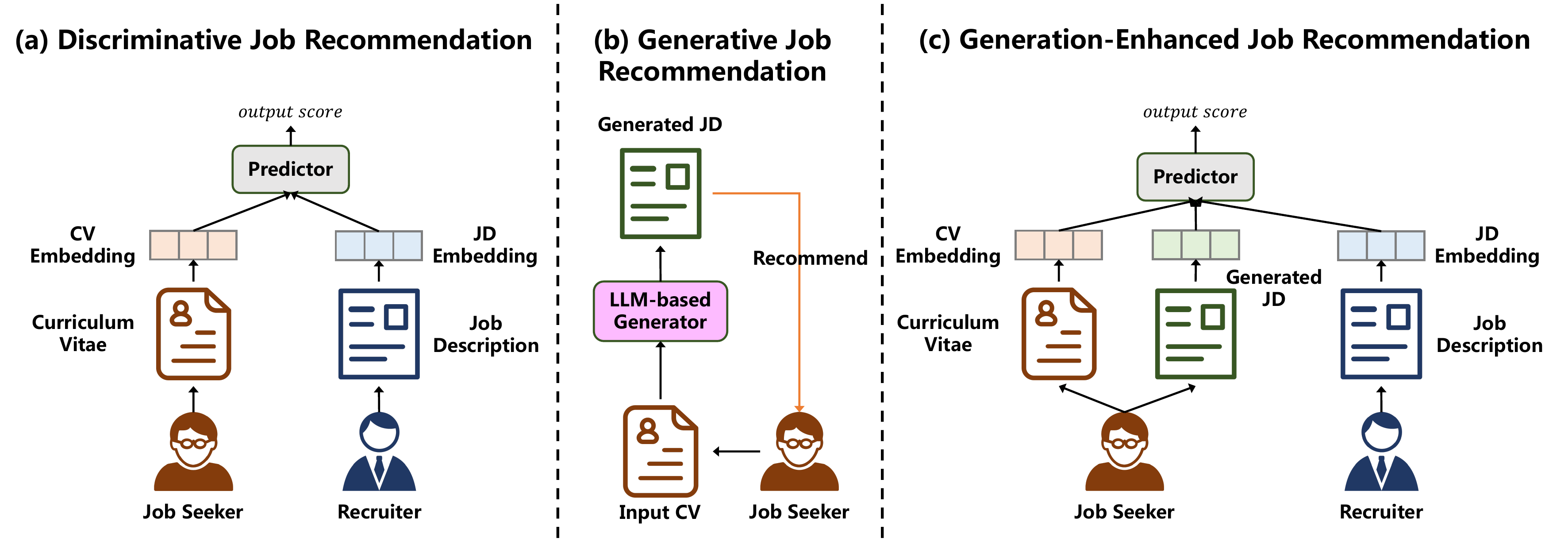}
\caption{Schematic diagram of three distinct job recommendation paradigms.}
\label{fig:intro}
\end{figure*}

To address the aforementioned challenges, in this paper, inspired by the recent progress in the field of Large Language Models (LLM), we propose a novel user-centered \textbf{G}enerat\textbf{I}ve job \textbf{R}ecommendation paradigm based on \textbf{L}LM called \textbf{GIRL}. 
As shown in Figure~\ref{fig:intro}, different from traditional discriminative job recommendation methods which aim to predict a matching score given a specific job seeker and job, GIRL aims to directly generate a personalized JD for a specific job seeker based on the remarkable generation ability of LLMs. 
We propose two ways to leverage the generated job descriptions. Firstly, the job descriptions generated by the LLMs represent the job that the model deems most suitable for the job seeker. 
Therefore, these descriptions can provide job seekers with references for their job seeking and career development planning. 
Meanwhile, it can improve the explainability of the whole recommender system.
Secondly, the generated results can be used to bridge the semantic gap between CVs and JDs, and further enhance the performance of traditional discriminative models.

However, it is non-trival to train an LLM for job recommendation. 
On one hand, given the significant differences between recommendation tasks and NLP tasks, the LLM needs to incorporate more domain-specific knowledge~\cite{wu2023survey}. 
On the other hand, to better assist with downstream recommendation tasks, the LLM needs to further learn from historical interaction records.
Therefore, inspired by the InstructGPT~\cite{ouyang2022training}, in this paper, we propose a three-step training methodology as:
\begin{enumerate}
    \item Supervised Fine-Tuning (SFT): This step aims to teach the LLM how to generate an appropriate JD based on a given CV. Specifically, we build a dataset consisting of previously matched CV-JD pairs, and use the intruction-tuning method to train the LLM generator.
    \item Reward Model Training (RMT): In this step, we build a dataset consists of matched and mismatched CV-JD pairs, which contains the recruiter feedback for the job seekers. 
    Then, we train a reward model to distinguish the matched CV-JD pairs from mismatched ones to mimic the real-world recruiter.
    \item Reinforcement Learning from Recruiter Feedback (RLRF): 
    In step three, we leverage Proximal Policy Optimization (PPO) based reinforcement learning method to further align the LLM to the recruiter preference captured by the reward model, making the LLM generation consider not only the preference of the job seeker but also the practical market demands.
\end{enumerate}
Finally, the major contribution of this article can be summarized as follows:
\begin{itemize}
    \item To the best of our knowledge, this is the first piece of work which proposes an LLM-based generative job recommendation paradigm.
    \item We propose a novel three-step training methodology with reinforcement learning from recruiter feedback to train a job description generator.
    \item We evaluated the quality of the generated results with the help of GhatGPT\footnote{https://chat.openai.com/}, and we further conducted extensive experiments on real-world dataset.
\end{itemize}

\section{RELATED WORK}
In this section, we will summarize the related works in the following three categories, respectively job recommendation, large language models, and LLMs for recommendation.
\subsection{Job Recommendation}
In the era of burgeoning online job platforms, a variety of novel job recommendation techniques have been introduced. These approaches can be primarily divided into two categories, respectively text-based methods and behavior-based methods. For text-based methods, PJFNN~\cite{DBLP:journals/tmis/ZhuZXMXDL18} 
formulated this task as a joint representation learning problem and utilized CNN-based models to get the representation of job seekers and recruiters, while APJFNN~\cite{DBLP:conf/sigir/QinZXZJCX18} enhanced the above model by taking the abilities of job seekers into consideration and used attention mechanisms for hierarchical ability-aware representation. IPJF~\cite{DBLP:conf/cikm/LeHSZ0019} conceived an interpretable model to match job seekers and recruiters in a multi-task learning framework.
For behavior-based methods, DPGNN~\cite{DBLP:conf/recsys/YangHSZWZ22} proposed to build an interaction graph between job seekers and recruiters to model the directed interactions. DPJF-MBS~\cite{DBLP:conf/dasfaa/FuL0SZW21} proposed to utilize memory networks to get the representation of the multi-behavior sequences of different job seekers and recruiters.

\subsection{Large Language Models}
Large Language Models (LLMs) are language models consisting of a neural network with many parameters (tens of millions to even trillions), and trained on large quantities of unlabeled text using self-supervised learning or semi-supervised learning methods~\cite{min2021recent,zhao2023survey}. 
Large language models primarily rely on the Transformer~\cite{vaswani2017attention} architecture, which has become the standard deep learning technique for Natural Language Processing (NLP). 
Existing LLMs can primarily be divided into two categories, respectively discriminative LLMs and generative LLMs.
For discriminative LLMs, BERT~\cite{DBLP:conf/naacl/DevlinCLT19} proposed a deep bidirectional transformer architecture, and further proposed a Masked Language Model (MLM) objective for model pre-training. 
Roberta~\cite{DBLP:journals/corr/abs-1907-11692} further refined the training process of BERT and achiever better performance.
XLNet~\cite{DBLP:conf/nips/YangDYCSL19} leveraged the permutation of the sequence order, enabling it to learn the context of a word based on all the words before and after it in a sentence.
For generative LLMs, GPT~\cite{radford2018improving} proposed to improve language understanding by generative pre-training. During pre-training, the model learns to predict the next word in a sentence, without any specific task in mind. 
GPT-2~\cite{radford2019language} and GPT-3~\cite{brown2020language} further increased the model scale and achieved better performance. 
InstructGPT~\cite{DBLP:conf/nips/Ouyang0JAWMZASR22} further proposed to fine-tune the GPT model using reinforcement learning from human feedback. 
Inspired by the above studies, in this paper, we also use reinforcement learning to fine-tune the JD generator and we use BERT for text embedding.

\subsection{LLMs for Recommendation}
LLMs have recently gained significant attention in the domain of recommendation systems~\cite{wu2023survey}.
Generally, existing studies can be divided into two categories, respectively recommendation based on discriminative LLMs and generative LLMs.
For the former category, U-BERT~\cite{qiu2021u} proposed a novel pre-training and fine-tuning method to leverage BERT for recommendation tasks. BERT4Rec~\cite{sun2019bert4rec} proposed to utilize BERT-based deep bidirectional self-attention architecture to model user behavior sequences.
For the latter category, some studies focus on utilizing the zero/few shot abilities of LLMs, and use the LLMs for recommendation by prompting without fine-tuning~\cite{liu2023chatgpt,hou2023large,sileo2022zero}. 
Moreover, some studies further fine-tune the LLMs, endeavoring to achieve better performance. 
For example, TALLRec~\cite{bao2023tallrec} proposed to fine-tuned the LLMs by recommendation tuning, where the input is the historical sequence of users and the output is the "yes or no" feedback. InstructRec~\cite{zhang2023recommendation} designed 39 instruction templates and automatically generated a large amount of instruction data for instruction tuning.
\section{Problem Definition}
Here we introduce the problem formulation of generative job recommendation and generation-enhanced job recommendation. 
Let \(\mathcal{S}\) and  \(\mathcal{J}\) denote the entire job seeker set and job set.
The feedback matrix between the job seekers and recruiters is denoted as $\mathcal{Z}\in \mathbb{R}^{N_s\times N_j}$, where $z_{s,j}=1$ means both job seeker $s$ and the recruiter of job $j$ are satisfied with each other and this pair is \textit{matched}, and $z_{s,j}=0$ means this pair is \textit{mismatched}.
$N_s$ and $N_j$ denote the numbers of the job seekers and jobs, respectively. 
Furthermore, each job seeker $s$ has a corresponding CV which can be formatted as $C=[w_{1},\dots,w_{l_s}]$, where $w_{i}$ is the $i$-th word in $C$ and $l_s$ is the length of $C$. 
Similarly, each job $j$ has a corresponding JD which can be formatted as $J=[v_{1},\dots,v_{l_j}]$, where $v_{i}$ is the $i$-th word in $J$ and $l_j$ is the length of $J$. Note that we omit some subscripts to facilitate the reading.

Traditionally, the objective of discriminative job recommendation is to train a scoring model that can compute the matching score between a given job seeker $s$ and a job $j$. 
However, this traditional paradigm can only recommend existing jobs for job seekers, which may not fulfill the needs of some job seekers. 
Therefore, in this paper, we propose a novel generative job recommendation paradigm which can be formulated as:
\begin{myDef}[Generative Job Recommendation]
Given a job seeker $s$ with the corresponding $C$, the goal of generative job recommendation is to train a generator \(\mathcal { G }\), which can generate a suitable JD for this user, i.e., $\mathcal { G }:C \to J^{\prime}$.
\end{myDef}

In the aforementioned definitions, the generated ${J}^{\prime}$  should has high quality and encompassing the most suitable job information for job seeker $s$, thereby providing meaningful guidance for $s$. Furthermore, in this paper, we propose that $J^{\prime}$ can also serve as a synopsis of job seeker $s$, contributing auxiliary support to traditional recommendation tasks. Along this line, the generation-enhanced job recommendation can be formulated as:
\begin{myDef}[Generation-Enhanced Job Recommendation]
Given a job seeker $s$ with the corresponding $C$, a job $j$ with the corresponding $J$, and the generated ${J}^{\prime}$, the goal of generation-enhanced job recommendation is to train a model $\mathcal{M}$, which can calculate the matching score between $s$ and $j$, i.e., $\mathcal {M}:C,J,J^{\prime}\to \mathbb{R}$.
\end{myDef}

\section{Generative Recommendation Framework}
\label{sec: GenRec}
\begin{figure*}
\centering
\includegraphics[width=\textwidth]{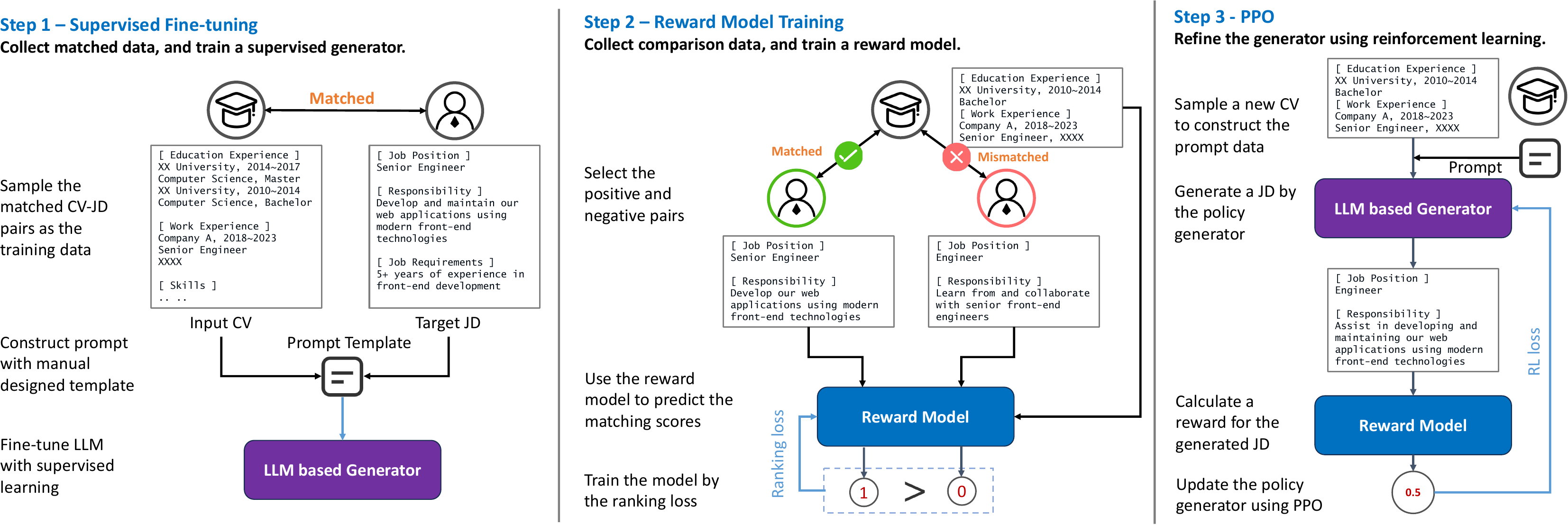}
\label{fig:training}
\caption{The training framework of the generative recommendation model.}
\label{fig:framework}
\end{figure*}

As shown in Figure~\ref{fig:training}, the generative recommendation framework is based on a large language model and consists of three training steps.
Specifically, we first convert the JD recommendation task to the NLG format with the manually designed prompt template, and utilize supervised fine-tuning to make the LLM generator understand the recommendation task. Second, we train a reward model to learn the recruiter feedback and capture the interaction information. Third, we utilize reinforcement learning to further align the generator with the recruiting market.
We will address the details of all steps in the following sub-sections.

\subsection{Supervised Fine-tuning}
In this training step, we propose to train the generator in the supervised fine-tuning way based on the matched CV-JD pairs. 
First, given a specific job seeker $s$ with the CV $C$ and a job $j$ with the JD $J$, we first build a prompt $T$ to describe the generation task as shown in Figure~\ref{fig:prompt}.
To maintain consistency with the training data, the original prompt is in Chinese. However, for better illustration, we have translated it to English in Figure~\ref{fig:prompt}.
The prompt template consists of the following four parts: 
\begin{figure}
\centering
\includegraphics[width=\linewidth]{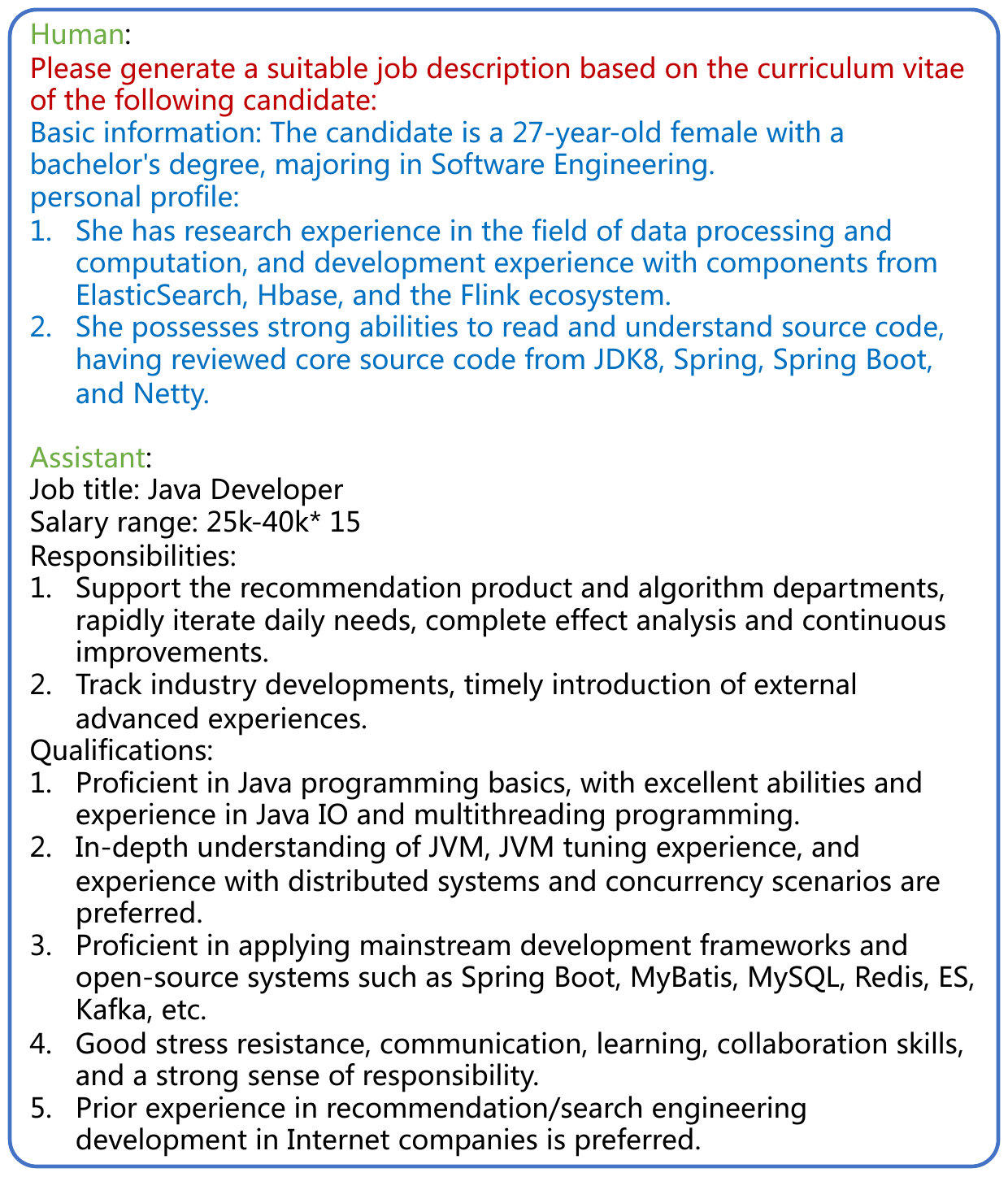}
\caption{The prompt template of training step one.}
\label{fig:prompt}
\end{figure}

\begin{itemize}
    \item Role: the green words, which aims to keep consistence with the instruction-tuning data of the our used backbone.
    \item Instruction: the black words, which describes the generation task via the human natural language.
    \item Input: the blue words, which contains the information of the job seeker.
    \item Output: the black words, which is the generation target, i.e., the JD text. Note that this part will be blank in the inference phase.
\end{itemize}

Then, we propose to train the generator with the casual language model pre-training task. Specifically, given the generator $\mathcal{G}$, the CV $C$, and the prompt template $T$, we optimize the negative log-likelihood for generating the JD $J$ as:

\begin{equation}
\begin{split}
    \mathcal{L}_{sft} &= - \log \Pr(C|J,T,\mathcal{G}) \\
                &= -\sum_{i=1}^{|l_j|} \log \Pr(v_{i}|v_{<i},C,T,\mathcal{G}),
\end{split}
\end{equation}
where $l_j$ is the length of $J$, $v_i$ is the $i$-th word in $J$. $\Pr(C|J,T,\mathcal{G})$ denotes the generation probability for $J$ of the generator model $\mathcal{G}$ given the job seeker feature $C$ and the prompt template $T$.


\subsection{Reward Model Training}
In this training step, our aim is to train a reward model $\mathcal{U}$ that can predict the matching score between a CV-JD pair, i.e., $\mathcal{U}: (C, J) \to \mathbb{R}$. The architecture of $\mathcal{U}$ is similar to that of the generator model $\mathcal{G}$, but it has a linear prediction head that outputs scalar values. Additionally, the parameter scale of $\mathcal{U}$ is smaller than that of $\mathcal{G}$.

To train the reward model $\mathcal{U}$, we collect pairwise training data and construct a ranking task. Typically, a job seeker applies for multiple jobs simultaneously and receives different feedback (matched or rejected) from recruiters. Therefore, we select a matched job $J^+$ and a mismatched job $J^-$ for each CV $C$ to construct comparable pairs. We then optimize the pairwise ranking loss to train $\mathcal{U}$ as follows:
\begin{equation}
    \mathcal{L}_{rmt} = \log \sigma (\mathcal{U}(C, J^+) - \mathcal{U}(C, J^-)),
\end{equation}
where $\sigma$ denotes the Sigmoid activation function.

This approach enables the reward model to capture the market preferences for job seekers based on the feedback from recruiters. Moreover, we can use the reward model to predict the matching score between a job seeker and a generated job description, thereby verifying the suitability of the recommendation in advance.

\subsection{Reinforcement Learning}

In this stage, we aim to improve the alignment between the generator $\mathcal{G}$ and the recruiter feedback acquired by the reward model $\mathcal{U}$ through reinforcement learning. 
Drawing inspiration from InstructGPT~\cite{ouyang2022training}, we employ the Proximal Policy Optimization (PPO)~\cite{DBLP:journals/corr/SchulmanWDRK17} algorithm to facilitate this alignment process. 
Specifically, we first utilize the generator $\mathcal{G}$ and the reward model $\mathcal{U}$ obtained from the first two training steps to initialize the actor-critic model, comprising the actor model $\mathcal{G}^a$ and critic model $\mathcal{U}^c$.
Next, we collect a RL training dataset, which only consists of the CVs of job seekers which do not appear in the first two stages.
Then, we use the PPO algorithm to train the actor-critic model based on these CVs while freezing the generator and the reward model.
Finally, we use the actor as the new generator model. The entire optimization algorithm is an iterative process and the ensuing sub-sections expound on the details of an iteration.
\subsubsection{Job Description Generation}
We first samples some CVs $\mathcal{C}^{r}$ from the training data and then leverage the actor model $\mathcal{G}^a$ to generate JDs $\mathcal{J}^{r}=\{\mathcal{G}^a(C)| C \in \mathcal{C}^{r}\}$ for these samples.
For simplicity, we take the $i$-th sample $\mathcal{C}^{r}_i$ with its corresponding generated JD $\mathcal{J}^{r}_i$ as the example to illustrate the following calculation steps.

\subsubsection{KL Divergence Computation}
To ensure the convergence and stability of the RL algorithm, the PPO algorithm uses KL divergence to limit the range of changes in the policy during each update.
The KL divergence is a metric for measuring the difference between the current policy, i.e., the actor model $\mathcal{G}^a$, and the old policy, i.e., $\mathcal{G}$. 

Specifically, given the pair of CV $\mathcal{C}^{r}_i$ and generated JD $\mathcal{J}^{r}_i$, we can estimate the KL divergence as follows:
\begin{equation}
\begin{split}
    &\textit{KL}(\mathcal{C}^{r}_i, \mathcal{J}^{r}_i) = \frac{1}{|\mathcal{J}^{r}_i|}\sum_{v_{j}\in \mathcal{J}^{r}_i} \left ( \textit{CE}(v_{i,j}) - 1 -\log \textit{CE}(v_{i,j}) \right ), \\
    &\textit{CE}(v_j) = \frac{\Pr(v_{j}|v_{i,<j},C,\mathcal{G}^a)}{\Pr(v_{j}|v_{i,<j},C,\mathcal{G})},
\end{split}
\end{equation}
where $v_{i,j}$ and $v_{i,<j}$ denote the $j$-th token and first $(j-1)$ tokens of the JD $\mathcal{J}^{r}_i$, respectively.

\subsubsection{Reward and Advantages Computation}
The final reward consists of two different parts, respectively the matching score predicted by the reward model and the KL divergence, and can be fomulated as follows:
\begin{equation}
    r_i = \mathcal{U}(\mathcal{C}^{r}_i, \mathcal{J}^{r}_i) - \lambda \textit{KL}(\mathcal{C}^{r}_i, \mathcal{J}^{r}_i),
\end{equation}
where $\lambda$ is the coefficient of the KL divergence.

Furthermore, the advantage value is the difference between the reward and the value of the input CV estimated by the critic model as:
\begin{equation}
    a_i = r_i - \mathcal{U}^c(\mathcal{C}^{r}_i, \_).
\end{equation}

\subsubsection{Actor Model Optimization}
After obtaining the above values, we can finally calculate the policy loss, i.e., the loss of actor model.
Here, we use the importance sampling and clip tricks to estimate the loss as:
\begin{equation}
     \mathcal{L}_{am} = \frac{1}{|\mathcal{J}^{r}_i|}\sum_{v_{j}\in \mathcal{J}^{r}_i}  \min \Big( \textit{CE}(v_{i,j})a_i, \operatorname{clip}(\textit{CE}(v_{i,j}))a_i \Big),
\end{equation}

\begin{equation}
      \operatorname{clip}(\textit{CE}(v_{i,j})) =
\begin{cases}
1 + \epsilon, & \textit{CE}(v_{i,j}) >  1 + \epsilon \\
\textit{CE}(v_{i,j}), &  1 - \epsilon  < \textit{CE}(v_{i,j}) <  1 + \epsilon \\
1 - \epsilon. &   \textit{CE}(v_{i,j}) <  1 - \epsilon
\end{cases}
\end{equation}

\subsubsection{Critic Model Optimization}
The critic model loss is the MSE loss between the reward value and the estimated state value as:
\begin{equation}
    \mathcal{L}_{cm} = (r_i - \mathcal{U}^c(\mathcal{C}^{r}_i, \_))^2
\end{equation}

The above five steps constitute one iteration of the optimization process.
Through minimizing the actor loss and critic loss, we can optimize two models.
In the RL process, the reward model and the generator model are froze.
Moreover, the whole RL process are shown in Algorithm~\ref{alg:ppo}.

\begin{algorithm}
\caption{Proximal Policy Optimization}
\label{alg:ppo}
\begin{algorithmic}[1]
\REQUIRE Initial actor model $\mathcal{G}^a$, critic model $\mathcal{U}^c$, optimization steps $K$, minibatch size $B_r$, epochs $E$, learning rates $\alpha_{am}$ and $\alpha_{cm}$, clipping parameter $\epsilon$,  KL coefficient $\lambda$.
\FOR{iteration = $1,2,\ldots$}
    \STATE Sample a set of CVs $\mathcal{C}^{r}$ from the training data.
    \STATE Generate JDs for the sampled CVs by the generator model $\mathcal{G}^a$, $\mathcal{J}^{r}$.
    \STATE Compute the discounted rewards $\{r_i\}_{i=1}^{i=|\mathcal{C}^{r}|}$ and the advantages $\{a_i\}_{i=1}^{i=|\mathcal{C}^{r}|}$ using Eq.(5) and (6).
    \STATE Update the actor model parameters $\theta(\mathcal{G}^a)$ and critic model parameters $\phi(\mathcal{U}^c)$ as follows:
        \FOR{epoch = $1,2,\ldots,E$}
            \STATE Shuffle the dataset $\mathcal{D}$.
            \STATE Divide $\mathcal{D}$ into minibatches of size $B_r$.
            \FOR{each minibatch}
                \STATE Compute the policy loss $\mathcal{L}_{am}$
                \STATE Compute the value function loss $\mathcal{L}_{cm}$
                \STATE Update the actor model parameters using the policy loss and learning rate $\alpha_{am}$:
                \begin{equation*}
                \theta(\mathcal{G}^a) \leftarrow \theta(\mathcal{G}^a) - \alpha_{am} \nabla_\theta \mathcal{L}_{am}
                \end{equation*}
                \STATE Update the critic model parameters using the value function loss and learning rate $\alpha_{cm}$:
                \begin{equation*}
                \phi(\mathcal{U}^c) \leftarrow \phi(\mathcal{U}^c) - \alpha_{cm} \nabla_\phi \mathcal{L}_{cm}
                \end{equation*}
            \ENDFOR
        \ENDFOR
\ENDFOR
\end{algorithmic}
\end{algorithm}

\section{Generation-Enhanced Recommendation Framework}
\label{sec: Enhanced Recommendation}
In Section~\ref{sec: GenRec} we introduced the paradigm of generative job recommendation. 
As we mentioned before, given a CV corresponding to a job seeker, we can utilize LLMs to generate the most suitable JD, thereby providing career development guidance for this job seeker. 
Furthermore, in this paper, we propose that we can actually regard the above paradigm as a feature extraction process, which can further enhance the performance of traditional discriminative recommendation methods. 
In this section, we delve into the details of how to leverage the generated results provided by LLMs for enhanced job recommendation.

\subsection{Basic Recommendation Model}
As shown in Figure~\ref{fig:intro} (a), in the paradigm of discriminative recommendation based on text matching, given a job seeker $s$ with the corresponding CV $C$, and a job $j$ with the corresponding JD $J$, we first need to get the text embedding based on a text encoder as:
\begin{equation}
    \mathbf{c} = Encoder(C),~~
    \mathbf{j} = Encoder(J).
\end{equation}
Then, we can get the matching score by feeding the above embedding vectors to a predictor. In this paper, we studied two different predictors, respectively MLP predictor as:
\begin{equation}
    score = MLP([\mathbf{c};\mathbf{j}]),
\end{equation}
where $[;]$ is the concatenation of two vectors, and dot predictor as follows:
\begin{equation}
    score = \mathbf{c} \cdot \mathbf{j},
\end{equation}
where $\cdot$ calculates the dot product of two vectors.

\subsection{Enhanced Recommendation Model}
As shown in Figure~\ref{fig:intro} (c), in the paradigm of generation-enhanced job recommendation, we can get the generated JD $J^{\prime}$ based on the CV $C$ and the LLM-based generator $\mathcal{G}$. 
Then, we can also get the text embedding of $J^{\prime}$ as:
\begin{equation}
\label{eq:jd embedding}
    \mathbf{j}^{\prime} = Encoder(J^{\prime}).
\end{equation}
After that, we propose two different ways to utilize $\mathbf{j}^{\prime}$ for enhancing the recommendation task corresponding to different predictor. Specifically, for the MLP predictor, we propose to calculate the matching score as:
\begin{equation}
\label{eq:score calculation}
    score = MLP([\mathbf{c};\mathbf{j};\mathbf{j}^{\prime}]).
\end{equation}
For the dot predictor, we first get the enhanced job seeker embedding as:
\begin{equation}
    \mathbf{c}^{\prime} = MLP([\mathbf{c};\mathbf{j}^{\prime}]).
\end{equation}
Then, we can calculate the dot product as:
\begin{equation}
    score = \mathbf{c}^{\prime} \cdot \mathbf{j}.
\end{equation}
\section{Experiments}
In this section, we first describe the dataset used in this paper. 
Then, we propose to evaluate our approch from two different perspectives. 
We further present some discussions and case studies on generative job recommendation. 
The experiments are mainly designed to answer the research questions as follows:
\begin{itemize}
    \item \textbf{RQ1}: Can our LLM-based generator generate high-quality JDs?
    \item \textbf{RQ2}: Can the generated results enhance the performance of discriminative job recommendation?
    \item \textbf{RQ3}: Whether the specially designed training methods for the LLM effective?
    \item \textbf{RQ4}: How do different settings influence the effectiveness of our model?
\end{itemize}

\subsection{Data Description and Preprocessing}

\begin{table}[tb]
  \small
  \caption{Statistics of the datasets.}
  \label{tab:statistics}   
  \centering
  \resizebox{\linewidth}{!}{
  \begin{tabular}{c|c}
  \hline
  Description & Number \\\hline
  \# of data for supervised fine-tuning&153,006\\
  \# of data for reward model training&303,929\\
  \# of data for reinforcement learning&37,600\\
  \# of data in training set for enhanced recommendation&37,158\\
  \# of data in validation set for enhanced recommendation&4,542\\
  \# of data in test set for enhanced recommendation&6,300\\
  \hline
  \end{tabular}
  }
\end{table}
The real-world datasets used in this paper comes from one of the largest online recruitment platform in China. 
In our datasets, each job seeker and recruiter is de-linked from the production system by securely hashing with one-time salt mapping. 
In this platform, each job seeker has a Curriculum Vitae (CV), encompassing their basic demographic information, educational background, and work experience among other details. 
Meanwhile, each job is associated with a Job Description (JD), detailing the responsibilities of the role, the compensation package, and so on. 
A variety of interaction types may occur between job seekers and jobs, such as browsing, applying, and matched. 
In this paper, we categorize these interactions into two major types, respectively matched and mismatched.

To train a large language model for generative job recommendation, we built the following three dataset:
\begin{itemize}
    \item Supervised Fine-tuning Dataset: This dataset contains multiple matched CV-JD pairs, ranging from Apr. 1, 2023, to Apr. 30, 2023.
    \item Reward Model Training Dataset: This dataset contains multiple matched and mismatched CV-JD pairs, ranging from May. 1, 2023 to May. 7, 2023.
    \item Reinforcement Learning Dataset: This dataset contains CVs only, ranging from May. 8, 2023, to May. 10, 2023.
\end{itemize}
Furthermore, to evaluate whether the generated results can enhance the performance of traditional discriminative models, we built the following dataset:
\begin{itemize}
    \item Enhanced Recommendation Dataset: This dataset contains multiple matched and mismatched CV-JD pairs, ranging from May. 8, 2023 to May. 31, 2023.
\end{itemize}
Detailed statistics of the above datasets are shown in Table~\ref{tab:statistics}.

\subsection{Evaluation and Baselines}
In this paper, we propose to evaluate the effectiveness of our GIRL approach from the following two perspectives. 
Firstly, with the assistance of ChatGPT, we evaluated the quality of the generated results from semantic perspective. 
Secondly, we evaluated whether the generated results can enhance the performance of discriminative recommendation.

\begin{figure}
\centering
\includegraphics[width=\linewidth]{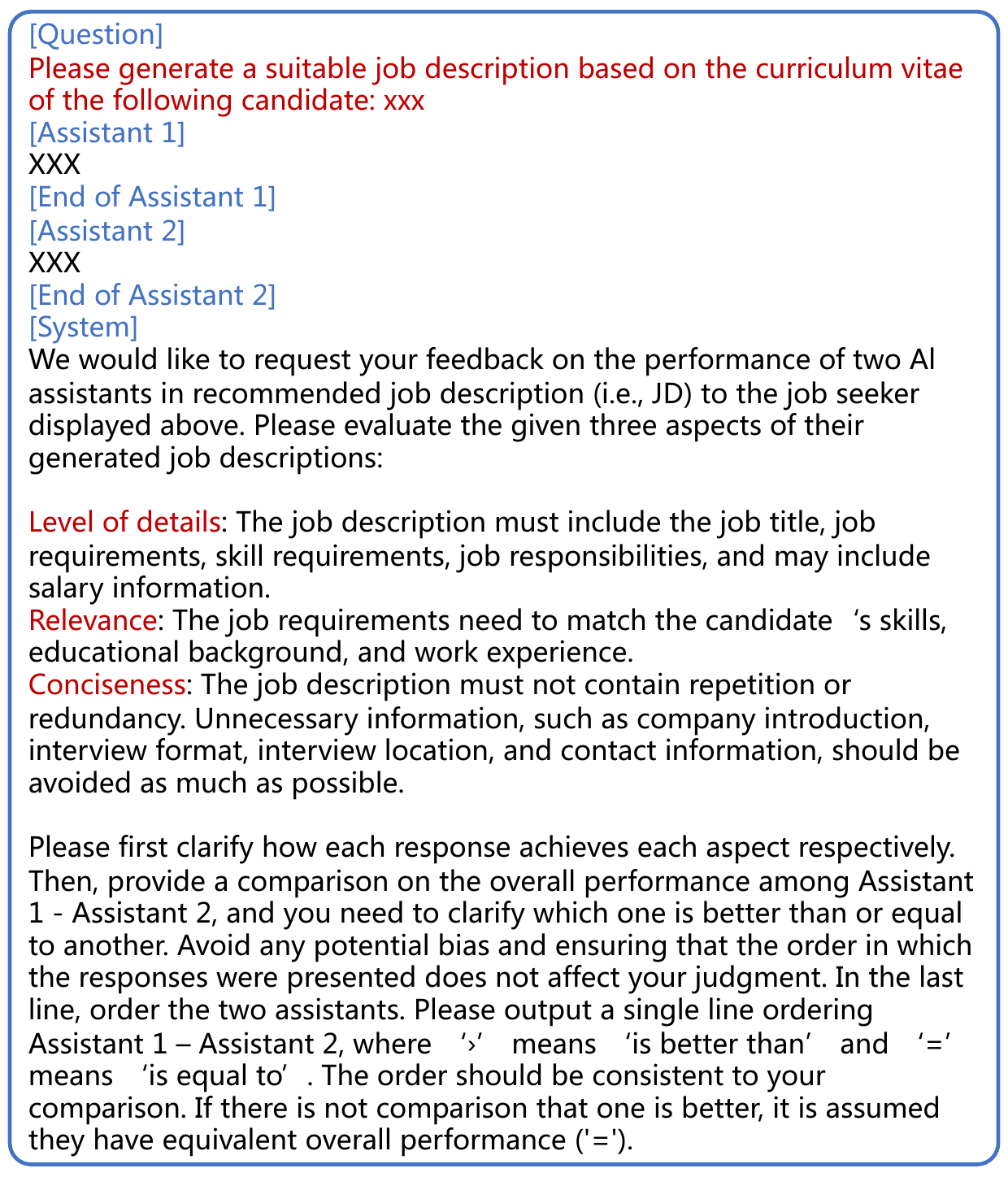}
\caption{The prompt template for generation quality evaluation.}
\label{fig:quality evaluation prompt}
\end{figure}

For generation quality evaluation, we first selected several baseline methods to compare with our method as:
\begin{itemize}
    \item GIRL: This is the method proposed in this paper which utilized both SFT and RL for fine-tuning.
    \item GIRL-SFT: This method is a simplified variant GIRL which only utilized SFT for fine-tuning.
    \item Other LLMs: BELLE-7b~\cite{BELLE}, BLOOMZ-7b~\cite{muennighoff2022crosslingual}, LLAMA-7b~\cite{touvron2023llama}.
\end{itemize}
Furthermore, we propose to utilize ChatGPT as the evaluator to compare the generation quality of these methods. 
Specifically, we first input the CV and two different JDs generated by two different methods into the prompt. 
We then request ChatGPT to evaluate the results from the following three different perspectives:
\begin{itemize}
    \item Level of details: Whether the generated JD contains enough necessary information about the job.
    \item Relevance: Whether the generated JD is suitable for the job seeker.
    \item Conciseness: Whether the generated JD is fluid and has high readability.
\end{itemize}
The detailed prompt template~\cite{llm-zoo-2023,phoenix-2023} for generation quality is shown in Figure~\ref{fig:quality evaluation prompt}, from which we can find that the output results of ChatGPT can be divided into three categories, respectively ``Win", ``Tie", and ``Lose". Based on the output results, given the dataset for generation quality evaluation, we selected ``Win Rate (Win)", ``Tie Rate (Tie)", and ``Lose Rate (Lose)", which is obtained by calculating the proportion of the above three results, as three different evaluation metrics. 
Note that we use boot strapping~\cite{hou2023large} strategy to avoid the position bias when using ChatGPT as the ranker. 
Furthermore, we define ``Advantage (Adv.)", which is the difference between ``Win Rate" and ``Lose Rate", as another evaluation metrics to reflect the relative improvement.

For evaluating the effectiveness of the generated results for recommendation enhancement, we selected several baseline methods to compare with our method as:
\begin{itemize}
    \item Base: This method is a traditional two-tower text matching model as shown in Figure~\ref{fig:intro} (a). We chose BERT~\cite{DBLP:conf/naacl/DevlinCLT19} as the text encoder for getting the CV and JD embedding.
    \item GIRL-SFT: As shown in Figure~\ref{fig:intro} (c), this method uses the generated JDs for recommendation enhancement. Only SFT is used for fine-tuning the LLM.
    \item GIRL: This method uses the generated JDs for recommendation enhancement. Both SFT and RL are used for fine-tuning the LLM.
\end{itemize}
Note that as we mentioned is Section~\ref{sec: Enhanced Recommendation}, we proposed two different methods for the predictor, respectively MLP and Dot. We will test the performance of different models with these two different predictors. 
We selected AUC and LogLoss as the evaluation metric for the enhanced recommendation task.

\subsection{Performance of Generation Quality (RQ1,RQ3)}

\begin{table}[tb]
  \small
  \caption{Comparison of Generation Quality Across Different Models.}
  \label{tab:Generation Quality}   
  \centering
  \resizebox{\linewidth}{!}{
  \begin{tabular}{l|c|c|c|c}
  \toprule
  Model Pair & Win & Tie & Lose & Adv. \\\midrule
  GIRL v.s. LLaMA-7b & 0.63 & 0.07 & 0.30 & 0.33 \\
  GIRL v.s. BLOOMZ-7b & 0.74 & 0.08 & 0.18 & 0.56 \\
  GIRL v.s. GIRL-SFT & 0.45 & 0.25 & 0.26 & 0.19 \\
  GIRL v.s. BELLE-7b & 0.45 & 0.17 & 0.36 & 0.09 \\
  \midrule
  GIRL-SFT v.s. LLaMA-7b & 0.55 & 0.03 & 0.42 & 0.13 \\
  GIRL-SFT v.s. BLOOMZ-7b & 0.73 & 0.07 & 0.19 & 0.54 \\
  GIRL-SFT v.s. BELLE-7b & 0.49 & 0.06 & 0.41 & 0.08 \\
  \midrule
  BELLE-7b v.s. LLaMA-7b & 0.61 & 0.04 & 0.34 & 0.27 \\
  BELLE-7b v.s. BLOOMZ-7b & 0.65 & 0.07 & 0.17 & 0.48 \\
  \bottomrule
  \end{tabular}
  }
\end{table}

To validate the quality of the JDs generated by our model, we first built a evaluation set with 200 different CVs which do not appear in other dataset. Then, we compared GIRL with all the baseline models on this dataset, and the results are shown in Table~\ref{tab:Generation Quality}. From the results, we can get the following observations:
\begin{enumerate}
    \item The performance of the BELLE model significantly surpasses that of LLaMA and BLOOMZ. This underlines that instruction-tuning with instructions on Chinese datasets can substantially enhance the quality of the outputs in Chinese.
    \item Both GIRL and GIRL-SFT outperform all the baseline methods, emphasizing the necessity of instruction tuning on domain-specific data.
    \item GIRL exceeds GIRL-SFT in performance, demonstrating that reinforcement learning can better align the results generated by the LLMs with human preferences, thereby improving the quality of generated results.
\end{enumerate}

\subsection{Performance of Enhanced Recommendaion (RQ2,RQ3)}
\label{sec:recommendaion result}
\begin{table}[tb]
  \small
  \caption{Overall Performance of Different Models on the Discriminative Job Recommendation.}
  \label{tab:enhanced}   
  \centering
  \resizebox{\linewidth}{!}{
  \begin{tabular}{c|c|c|c}
  \toprule
  Predictor& Model & AUC($\uparrow$)&LogLoss($\downarrow$) \\\midrule
  \multirow{3}{*}{MLP}&Base&0.6349&0.4043\\
  &GIRL-SFT&0.6438(+1.4\%)&0.3973(+1.7\%)\\
  &GIRL&\textbf{0.6476}(+2.0\%)&\textbf{0.3908}(+3.3\%)\\\midrule
  \multirow{3}{*}{Dot}&Base&0.6258&0.4964\\
  &GIRL-SFT&0.6291(+0.5\%)&0.3688(+20.3\%)\\
  &GIRL&\textbf{0.6436}(+2.8\%)&\textbf{0.3567}(+28.1\%)\\
  
  \bottomrule
  \end{tabular}
  }
\end{table}

To demonstrate the effectiveness of the generation results for enhancing the discriminative recommendation task, we compare GIRL with all the baseline methods, and the results are shown in Table~\ref{tab:enhanced}. From the results, we can get the following observations:
\begin{enumerate}
    \item Both GIRL and GIRL-SFT outperform the Base model, demonstrating that the JDs generated by fine-tuned LLMs can effectively enhance the performance of discriminative job recommendation.
    \item GIRL surpasses GIRL-SFT on all the evaluation metrics. The rationale behind this is that through the reward model training stage, our reward model encapsulates extensive real-world experiences. By incorporating this knowledge into the LLMs through reinforcement learning, the generated JDs are enabled to  capture job-seeker traits precisely and align with the preferences of recruiters better.
\end{enumerate}

\subsection{Discussion on Generation Number (RQ4)}
\begin{figure}
\centering
\includegraphics[width=\linewidth]{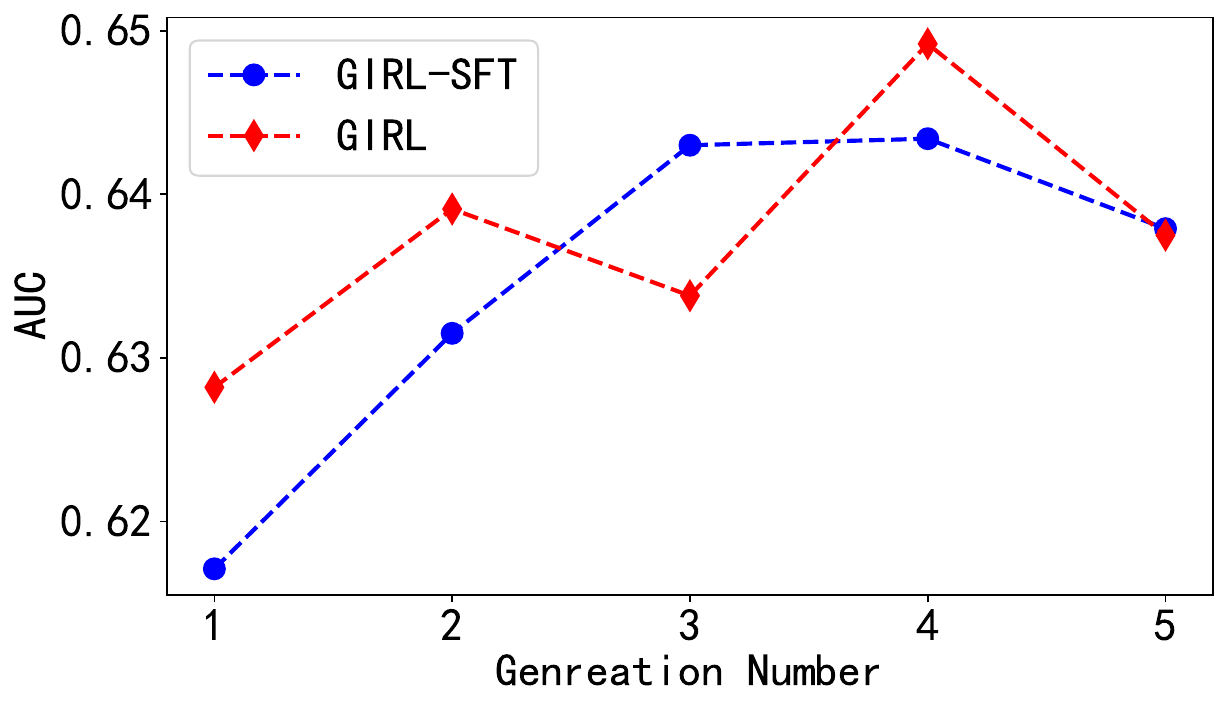}
\caption{Performance of different models with different generation number.}
\label{fig:gen num}
\end{figure}

In Section~\ref{sec: Enhanced Recommendation} we studied how to utilize the generated JD for enhancing discriminative recommendation, where we focus on utilizing a single JD. 
Indeed, owing to the inherent randomness in the text generation process, given a specific CV, the LLM is capable of generating multiple distinct JDs. 
In this section, we will explore how to utilize multiple generated JDs and discuss the influence of the number of JDs on the model performance. 
Specifically, given multiple JDs, we first get the text embedding of each JD by Equation~\ref{eq:jd embedding}. 
Then, we use mean pooling to fuse these JD embedding, and calculate the matching score following Equation~\ref{eq:score calculation}. 
The results are shown in Figure~\ref{fig:gen num}. 
Note that we employed only 75\% of the data in Section~\ref{sec:recommendaion result} to accelerate the computation process. From the results, we can find that as the number of generated JDs increases, the model performance initially improves before subsequently declining. This suggests that moderately increasing the number of generated JDs can further enhance model performance. However, a larger number of JDs also implies a substantial increase in computational cost. Moreover, the performance of GIRL surpasses that of GIRL-SFT in most cases, which once again affirms the superiority of the RL-based fine-tuning method proposed in this paper.

\begin{figure*}
\centering
\includegraphics[width=\textwidth]{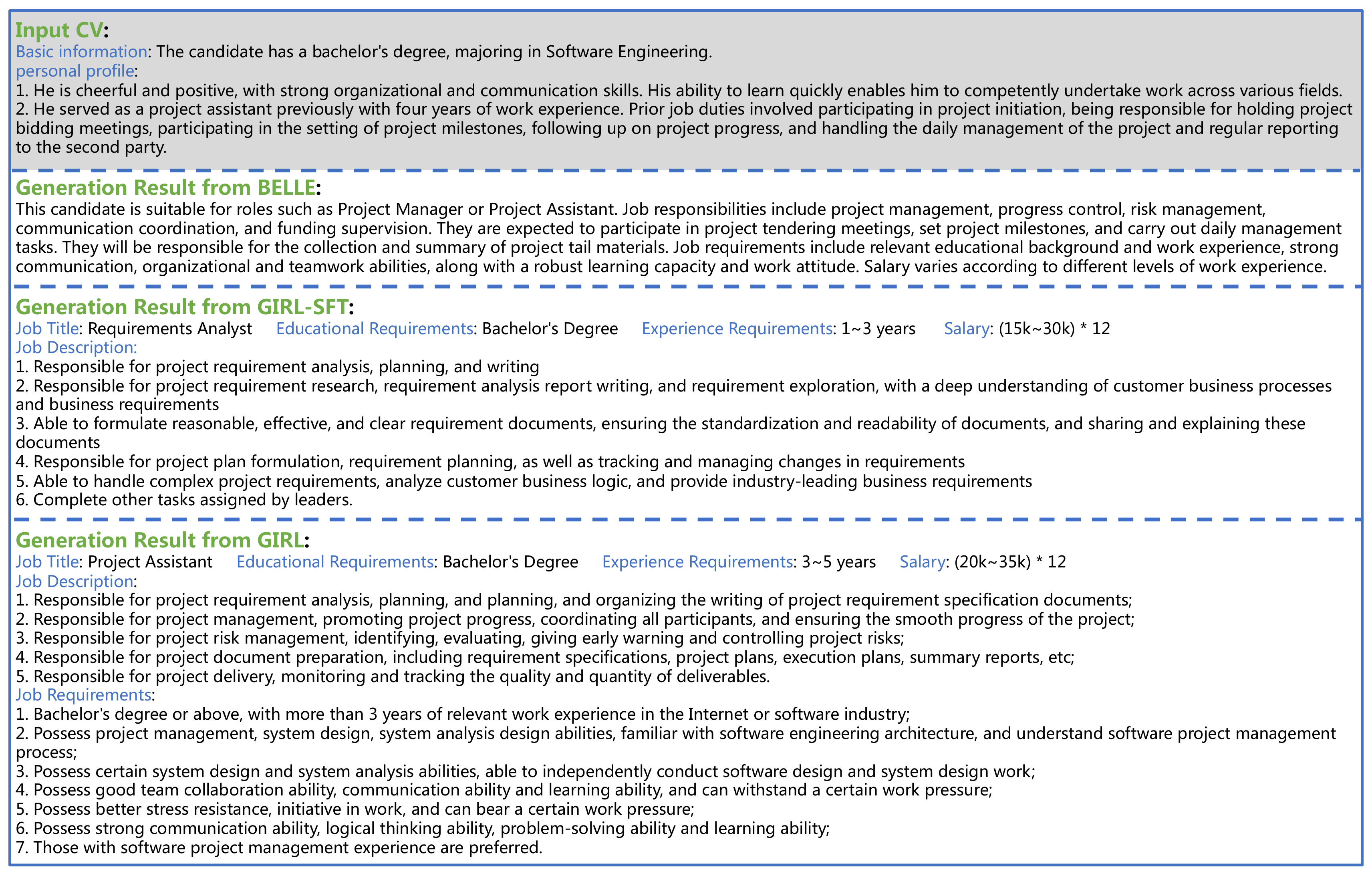}
\caption{A case study of the generation results from different models.}
\label{fig:case}
\end{figure*}

\subsection{Discussion on Cold Start (RQ4)}
\begin{table}[tb]
  \small
  \caption{Performance of Different Models on the Discriminative Job Recommendation Under Cold-start Condition.}
  \label{tab:cold start}   
  \centering
  \resizebox{\linewidth}{!}{
  \begin{tabular}{c|c|c|c}
  \toprule
  Predictor& Model & AUC($\uparrow$)&LogLoss($\downarrow$) \\\midrule
  \multirow{3}{*}{MLP}&Base&0.6198&0.4270\\
  &GIRL-SFT&0.6293(+1.5\%)&\textbf{0.4154}(+2.8\%)\\
  &GIRL&\textbf{0.6347}(+2.4\%)&0.4229(+1.0\%)\\\midrule
  \multirow{3}{*}{Dot}&Base&0.6136&0.5233\\
  &GIRL-SFT&0.6231(+1.5\%)&0.3827(+26.9\%)\\
  &GIRL&\textbf{0.6457}(+5.2\%)&\textbf{0.3673}(+29.8\%)\\
  
  \bottomrule
  \end{tabular}
  }
\end{table}

In this section, we will explore the performance of different models under cold-start condition on the  discriminative job recommendation task. 
Specifically, cold start condition refers to recommending jobs for job seekers who have not appeared in the training set. 
The results are shown in Table~\ref{tab:cold start}. 
Compared with Table~\ref{tab:enhanced}, we can find that the performance improvement of our models in cold-start conditions is more significant. 
This indicates that the JDs generated by LLMs can more effectively assist discriminative recommendation models in enhancing performance under cold-start conditions.

\subsection{Case Study}
In this section, we will conduct a case study of the generated results from different models for the same CV, and the results are shown in Figure~\ref{fig:case}. 
From the results we can find that the vanilla BELLE model without finetuning fails to generate job JDs in a standard format, and the generated JDs present vague descriptions of job-related skills and requirements, providing inadequate guidance for job seekers. Moreover, we can find that after being trained through reinforcement learning, the GIRL model generates results that are more standardized in format, more detailed and comprehensive in content, and more aligned with the individual circumstances of job seekers. The above results demonstrate the effectiveness of the three-stage training method proposed in this paper.

\section{CONCLUSION}
Reflecting on the recent advancements in the field of Large Language Models, this study presented a novel generative job recommendation paradigm named GeneratIve job Recommendation based on Large language model (GIRL).
Specifically, we first utilized supervised fine-tuning to guide the LLM-based generator in creating an appropriate job description given a specific curriculum vitae.
Subsequently, we developed a reward model predicated on feedback from recruiters, and then implemented a proximal policy optimization based reinforcement learning methodology to synchronize the generator with recruiter preferences.
Furthermore, we proposed to enhance the job seeker features by the generated results, aiming to improve the performance of the discriminative job recommendation model.
The series of experiments conducted on a real-world dataset from a large-scale online recruitment platform provided substantial evidence of the effectiveness of our proposed approach.

\bibliographystyle{IEEEtran}
\bibliography{main}

\begin{thebibliography}{10}
\providecommand{\url}[1]{#1}
\csname url@samestyle\endcsname
\providecommand{\newblock}{\relax}
\providecommand{\bibinfo}[2]{#2}
\providecommand{\BIBentrySTDinterwordspacing}{\spaceskip=0pt\relax}
\providecommand{\BIBentryALTinterwordstretchfactor}{4}
\providecommand{\BIBentryALTinterwordspacing}{\spaceskip=\fontdimen2\font plus
\BIBentryALTinterwordstretchfactor\fontdimen3\font minus
  \fontdimen4\font\relax}
\providecommand{\BIBforeignlanguage}[2]{{%
\expandafter\ifx\csname l@#1\endcsname\relax
\typeout{** WARNING: IEEEtran.bst: No hyphenation pattern has been}%
\typeout{** loaded for the language `#1'. Using the pattern for}%
\typeout{** the default language instead.}%
\else
\language=\csname l@#1\endcsname
\fi
#2}}
\providecommand{\BIBdecl}{\relax}
\BIBdecl

\bibitem{DBLP:conf/cikm/LeHSZ0019}
R.~Le, W.~Hu, Y.~Song, T.~Zhang, D.~Zhao, and R.~Yan, ``Towards effective and
  interpretable person-job fitting,'' in \emph{Proceedings of the 28th {ACM}
  International Conference on Information and Knowledge Management, {CIKM}
  2019, Beijing, China, November 3-7, 2019}.\hskip 1em plus 0.5em minus
  0.4em\relax {ACM}, 2019, pp. 1883--1892.

\bibitem{DBLP:conf/sigir/QinZXZJCX18}
C.~Qin, H.~Zhu, T.~Xu, C.~Zhu, L.~Jiang, E.~Chen, and H.~Xiong, ``Enhancing
  person-job fit for talent recruitment: An ability-aware neural network
  approach,'' in \emph{The 41st International {ACM} {SIGIR} Conference on
  Research {\&} Development in Information Retrieval, {SIGIR} 2018, Ann Arbor,
  MI, USA, July 08-12, 2018}.\hskip 1em plus 0.5em minus 0.4em\relax {ACM},
  2018, pp. 25--34.

\bibitem{DBLP:journals/tmis/ZhuZXMXDL18}
C.~Zhu, H.~Zhu, H.~Xiong, C.~Ma, F.~Xie, P.~Ding, and P.~Li, ``Person-job fit:
  Adapting the right talent for the right job with joint representation
  learning,'' \emph{{ACM} Trans. Manag. Inf. Syst.}, vol.~9, no.~3, pp.
  12:1--12:17, 2018.

\bibitem{wu2023survey}
L.~Wu, Z.~Zheng, Z.~Qiu, H.~Wang, H.~Gu, T.~Shen, C.~Qin, C.~Zhu, H.~Zhu,
  Q.~Liu \emph{et~al.}, ``A survey on large language models for
  recommendation,'' \emph{arXiv preprint arXiv:2305.19860}, 2023.

\bibitem{ouyang2022training}
L.~Ouyang, J.~Wu, X.~Jiang, D.~Almeida, C.~Wainwright, P.~Mishkin, C.~Zhang,
  S.~Agarwal, K.~Slama, A.~Ray \emph{et~al.}, ``Training language models to
  follow instructions with human feedback,'' \emph{Advances in Neural
  Information Processing Systems}, vol.~35, pp. 27\,730--27\,744, 2022.

\bibitem{DBLP:conf/recsys/YangHSZWZ22}
C.~Yang, Y.~Hou, Y.~Song, T.~Zhang, J.~Wen, and W.~X. Zhao, ``Modeling two-way
  selection preference for person-job fit,'' in \emph{RecSys '22: Sixteenth
  {ACM} Conference on Recommender Systems, Seattle, WA, USA, September 18 - 23,
  2022}.\hskip 1em plus 0.5em minus 0.4em\relax {ACM}, 2022, pp. 102--112.

\bibitem{DBLP:conf/dasfaa/FuL0SZW21}
B.~Fu, H.~Liu, Y.~Zhu, Y.~Song, T.~Zhang, and Z.~Wu, ``Beyond matching:
  Modeling two-sided multi-behavioral sequences for dynamic person-job fit,''
  in \emph{Database Systems for Advanced Applications - 26th International
  Conference, {DASFAA} 2021, Taipei, Taiwan, April 11-14, 2021, Proceedings,
  Part {II}}, ser. Lecture Notes in Computer Science, vol. 12682.\hskip 1em
  plus 0.5em minus 0.4em\relax Springer, 2021, pp. 359--375.

\bibitem{min2021recent}
B.~Min, H.~Ross, E.~Sulem, A.~P.~B. Veyseh, T.~H. Nguyen, O.~Sainz, E.~Agirre,
  I.~Heinz, and D.~Roth, ``Recent advances in natural language processing via
  large pre-trained language models: A survey,'' \emph{arXiv preprint
  arXiv:2111.01243}, 2021.

\bibitem{zhao2023survey}
W.~X. Zhao, K.~Zhou, J.~Li, T.~Tang, X.~Wang, Y.~Hou, Y.~Min, B.~Zhang,
  J.~Zhang, Z.~Dong \emph{et~al.}, ``A survey of large language models,''
  \emph{arXiv preprint arXiv:2303.18223}, 2023.

\bibitem{vaswani2017attention}
A.~Vaswani, N.~Shazeer, N.~Parmar, J.~Uszkoreit, L.~Jones, A.~N. Gomez,
  {\L}.~Kaiser, and I.~Polosukhin, ``Attention is all you need,''
  \emph{Advances in neural information processing systems}, vol.~30, 2017.

\bibitem{DBLP:conf/naacl/DevlinCLT19}
J.~Devlin, M.~Chang, K.~Lee, and K.~Toutanova, ``{BERT:} pre-training of deep
  bidirectional transformers for language understanding,'' in \emph{{NAACL-HLT}
  {(1)}}.\hskip 1em plus 0.5em minus 0.4em\relax Association for Computational
  Linguistics, 2019, pp. 4171--4186.

\bibitem{DBLP:journals/corr/abs-1907-11692}
Y.~Liu, M.~Ott, N.~Goyal, J.~Du, M.~Joshi, D.~Chen, O.~Levy, M.~Lewis,
  L.~Zettlemoyer, and V.~Stoyanov, ``Roberta: {A} robustly optimized {BERT}
  pretraining approach,'' \emph{CoRR}, vol. abs/1907.11692, 2019.

\bibitem{DBLP:conf/nips/YangDYCSL19}
Z.~Yang, Z.~Dai, Y.~Yang, J.~G. Carbonell, R.~Salakhutdinov, and Q.~V. Le,
  ``Xlnet: Generalized autoregressive pretraining for language understanding,''
  in \emph{NeurIPS}, 2019, pp. 5754--5764.

\bibitem{radford2018improving}
A.~Radford, K.~Narasimhan, T.~Salimans, I.~Sutskever \emph{et~al.}, ``Improving
  language understanding by generative pre-training,'' 2018.

\bibitem{radford2019language}
A.~Radford, J.~Wu, R.~Child, D.~Luan, D.~Amodei, I.~Sutskever \emph{et~al.},
  ``Language models are unsupervised multitask learners,'' \emph{OpenAI blog},
  vol.~1, no.~8, p.~9, 2019.

\bibitem{brown2020language}
T.~Brown, B.~Mann, N.~Ryder, M.~Subbiah, J.~D. Kaplan, P.~Dhariwal,
  A.~Neelakantan, P.~Shyam, G.~Sastry, A.~Askell \emph{et~al.}, ``Language
  models are few-shot learners,'' \emph{Advances in neural information
  processing systems}, vol.~33, pp. 1877--1901, 2020.

\bibitem{DBLP:conf/nips/Ouyang0JAWMZASR22}
L.~Ouyang, J.~Wu, X.~Jiang, D.~Almeida, C.~L. Wainwright, P.~Mishkin, C.~Zhang,
  S.~Agarwal, K.~Slama, A.~Ray, J.~Schulman, J.~Hilton, F.~Kelton, L.~Miller,
  M.~Simens, A.~Askell, P.~Welinder, P.~F. Christiano, J.~Leike, and R.~Lowe,
  ``Training language models to follow instructions with human feedback,'' in
  \emph{NeurIPS}, 2022.

\bibitem{qiu2021u}
Z.~Qiu, X.~Wu, J.~Gao, and W.~Fan, ``U-bert: Pre-training user representations
  for improved recommendation,'' in \emph{Proceedings of the AAAI Conference on
  Artificial Intelligence}, vol.~35, no.~5, 2021, pp. 4320--4327.

\bibitem{sun2019bert4rec}
F.~Sun, J.~Liu, J.~Wu, C.~Pei, X.~Lin, W.~Ou, and P.~Jiang, ``Bert4rec:
  Sequential recommendation with bidirectional encoder representations from
  transformer,'' in \emph{Proceedings of the 28th ACM international conference
  on information and knowledge management}, 2019, pp. 1441--1450.

\bibitem{liu2023chatgpt}
J.~Liu, C.~Liu, R.~Lv, K.~Zhou, and Y.~Zhang, ``Is chatgpt a good recommender?
  a preliminary study,'' \emph{arXiv preprint arXiv:2304.10149}, 2023.

\bibitem{hou2023large}
Y.~Hou, J.~Zhang, Z.~Lin, H.~Lu, R.~Xie, J.~McAuley, and W.~X. Zhao, ``Large
  language models are zero-shot rankers for recommender systems,'' \emph{arXiv
  preprint arXiv:2305.08845}, 2023.

\bibitem{sileo2022zero}
D.~Sileo, W.~Vossen, and R.~Raymaekers, ``Zero-shot recommendation as language
  modeling,'' in \emph{Advances in Information Retrieval: 44th European
  Conference on IR Research, ECIR 2022, Stavanger, Norway, April 10--14, 2022,
  Proceedings, Part II}.\hskip 1em plus 0.5em minus 0.4em\relax Springer, 2022,
  pp. 223--230.

\bibitem{bao2023tallrec}
K.~Bao, J.~Zhang, Y.~Zhang, W.~Wang, F.~Feng, and X.~He, ``Tallrec: An
  effective and efficient tuning framework to align large language model with
  recommendation,'' \emph{arXiv preprint arXiv:2305.00447}, 2023.

\bibitem{zhang2023recommendation}
J.~Zhang, R.~Xie, Y.~Hou, W.~X. Zhao, L.~Lin, and J.-R. Wen, ``Recommendation
  as instruction following: A large language model empowered recommendation
  approach,'' \emph{arXiv preprint arXiv:2305.07001}, 2023.

\bibitem{DBLP:journals/corr/SchulmanWDRK17}
J.~Schulman, F.~Wolski, P.~Dhariwal, A.~Radford, and O.~Klimov, ``Proximal
  policy optimization algorithms,'' \emph{CoRR}, vol. abs/1707.06347, 2017.

\bibitem{BELLE}
Y.~Ji, Y.~Deng, Y.~Gong, Y.~Peng, Q.~Niu, B.~Ma, and X.~Li, ``Belle: Be
  everyone's large language model engine,''
  \url{https://github.com/LianjiaTech/BELLE}, 2023.

\bibitem{muennighoff2022crosslingual}
N.~Muennighoff, T.~Wang, L.~Sutawika, A.~Roberts, S.~Biderman, T.~L. Scao,
  M.~S. Bari, S.~Shen, Z.-X. Yong, H.~Schoelkopf, X.~Tang, D.~Radev, A.~F. Aji,
  K.~Almubarak, S.~Albanie, Z.~Alyafeai, A.~Webson, E.~Raff, and C.~Raffel,
  ``Crosslingual generalization through multitask finetuning,'' 2022.

\bibitem{touvron2023llama}
H.~Touvron, T.~Lavril, G.~Izacard, X.~Martinet, M.-A. Lachaux, T.~Lacroix,
  B.~Rozi{\`e}re, N.~Goyal, E.~Hambro, F.~Azhar \emph{et~al.}, ``Llama: Open
  and efficient foundation language models,'' \emph{arXiv preprint
  arXiv:2302.13971}, 2023.

\bibitem{llm-zoo-2023}
Z.~Chen, J.~Chen, H.~Zhang, F.~Jiang, G.~Chen, F.~Yu, T.~Wang, J.~Liang,
  C.~Zhang, Z.~Zhang, J.~Li, X.~Wan, H.~Li, and B.~Wang, ``Llm zoo:
  democratizing chatgpt,'' \url{https://github.com/FreedomIntelligence/LLMZoo},
  2023.

\bibitem{phoenix-2023}
Z.~Chen, F.~Jiang, J.~Chen, T.~Wang, F.~Yu, G.~Chen, H.~Zhang, J.~Liang,
  C.~Zhang, Z.~Zhang, J.~Li, X.~Wan, B.~Wang, and H.~Li, ``Phoenix:
  Democratizing chatgpt across languages,'' \emph{arXiv preprint
  arXiv:2304.10453}, 2023.

\end{thebibliography}

\end{document}